\documentclass[11pt,leqno,fleqn]{article}
\usepackage[dvips]{graphicx}

\usepackage{algorithm}
\usepackage{algorithmic}

\newenvironment{ISItext}
{\normalsize\rm\setlength{\parindent}{1cm}\setlength{\parskip}{0pt}}
{\vskip 12pt}

\begin{document}
\newcommand{\ISItitle}[1]{\vskip 0pt\setlength{\parindent}{0cm}\Large\textbf{#1}\vskip 12pt}
\newcommand{\ISIsubtitleA}[1]{\normalsize\rm\setlength{\parindent}{0cm}\textbf{#1}\vskip 12pt}
\newcommand{\ISIsubtitleB}[1]{\normalsize\rm\setlength{\parindent}{0cm}\textbf{#1}\vskip 12pt}
\newcommand{\ISIsubtitleFig}[1]{\normalsize\rm\setlength{\parindent}{0cm}
\textbf{\textit{#1}}\vskip 12pt}
\newcommand{\ISIauthname}[1]{\normalsize\rm\setlength{\parindent}{0cm}#1 \\}
\newcommand{\ISIauthaddr}[1]{\normalsize\rm\setlength{\parindent}{0cm}\it #1 \vskip 12pt}

\ISItitle{Non-Euclidean statistical analysis of covariance matrices and diffusion tensors}

\ISIauthname{Dryden, Ian} 
\ISIauthaddr{University of South Carolina, Department of Statistics,\\
Columbia SC 29208, USA\\ 
E-mail: dryden@mailbox.sc.edu}

\ISIauthname{Koloydenko, Alexey} 
\ISIauthaddr{Royal Holloway University of London, Department of Mathematics,\\
Egham, Surrey, TW20 0EX, UK\\ 
E-mail: a.koloydenko@rhul.ac.uk}

\ISIauthname{Zhou, Diwei} 
\ISIauthaddr{University of Nottingham, School of Mathematical Sciences,\\
Nottingham, NG7 2RD, UK\\ 
E-mail: pmxdz@nottingham.ac.uk}

\ISIauthname{Li, Bai} 
\ISIauthaddr{University of Nottingham, School of Computer Science,\\
Nottingham, NG8 1BB, UK\\ 
E-mail: bai@cs.nott.ac.uk}

\begin{ISItext}

\section{Introduction}
The statistical analysis of covariance matrices occurs in many important 
applications, e.g. 
in diffusion tensor imaging  and
longitudinal data analysis. 
We consider the situation where it is of interest to estimate an average covariance matrix, 
describe its anisotropy, to carry out principal geodesic analysis  
and to interpolate between covariance matrices. 

An important difference with 
standard statistical techniques is that non-Euclidean distances are most natural 
for comparing covariance matrices, which are symmetric, semi-positive
definite matrices. 

\section{Diffusion tensors}
In medical image analysis a particular type of covariance matrix arises in 
diffusion weighted imaging called a diffusion tensor. The diffusion tensor 
is a $3 \times 3$ covariance matrix which is estimated at each voxel in 
the brain, and is obtained by fitting a physically-motivated model on 
measurements from the Fourier transform of the molecule displacement density
(Basser et al., 1994).  

In the diffusion tensor model the water molecules at a voxel diffuse   according to 
a multivariate normal model centred on the voxel and with 
covariance matrix $\Sigma$.  The displacement of a water molecule 
$x \in \mathbf R^3$ has probability density function 
$$f(x) = {\frac{1}{(2\pi)^{3/2} | \Sigma |^{1/2}} } 
\exp ( - \frac{1}{2} x^T \Sigma^{-1} x ) . $$ 
The convention is to call $D = \Sigma/2$ the diffusion tensor, 
which is a symmetric positive semi-definite matrix. The diffusion tensor is estimated at each 
voxel in the image from the available MR images. 
The MR scanner has a set of magnetic field gradients applied at directions 
$g_1,g_2,\ldots,g_m \in {R}P^2$ with scanner gradient parameter $b$, 
where ${R}P^2$ is the real projective space of axial directions (with $g_j \equiv  - g_j$, $\| g_j \| = 1$). 
The data at a voxel consist of signals  $(Z_0,Z_1,\ldots,Z_m)$
which are related to the Fourier 
transform of the displacement density in axial direction $g_j \in {R}P^2, \; j=1,\ldots,m$, and  the reading 
$Z_0$ is obtained with no gradient ($b=0$).   
The Fourier transform in axial direction 
$g \in {R}P^2$ of the multivariate Gaussian displacement density is given by 
$$ {\cal F}(g) = \int \exp( i \sqrt b g^T x) f(x) dx = 
\exp(-b g^T D g ) , $$
and the theoretical model for 
the signals is 
$$Z_j = Z_0 {\cal F}(g_j) = Z_0 \exp(-b g_j^T D g_j ), \; \; j=1,\ldots,m.$$  
There is a variety of methods available 
for estimating  $D$ from the data $(Z_0,Z_1,\ldots,Z_m)$ at each voxel 
(see Alexander, 2005), including least squares regression and Bayesian estimation 
(e.g. Zhou et al., 2008). 
Noise models include log-Gaussian, Gaussian and more recently 
Rician noise (e.g. Fillard et al., 2007).  
  A common method for visualizing a diffusion tensor is 
an ellipsoid with principal axes given by the eigenvectors of $D$, and lengths of axes 
proportional to $\sqrt{\lambda_i}, \; i=1,2,3.$ 



If a sample of diffusion tensors  
is available we may wish to estimate an average diffusion tensor matrix, 
investigate the structure of variability in diffusion tensors or 
interpolate at higher spatial resolution between two or more estimated diffusion 
tensor matrices.

A strongly anisotropic diffusion tensor indicates a
strong direction of white matter fibre tracts, and plots of measures of anisotropy are 
very useful to neurologists. 
A measure that is very commonly used in diffusion tensor imaging is Fractional Anisotropy
\begin{equation}
FA = \left\{ \frac{k}{k-1} \sum_{i=1}^k (\lambda_i - \bar \lambda)^2 / 
\sum_{i=1}^k \lambda_i^2 \right\}^{1/2}  \; \; , \label{FA}
\end{equation}
where $ 0 \le FA \le 1$ and $\lambda_i$ are the eigenvalues of the diffusion tensor matrix. 
Note that $FA \approx 1$ if $\lambda_1 >> \lambda_i, i > 1$ (very strong principal axis) and $FA = 0$ for isotropy. In diffusion tensor imaging $k = 3$.

\section{Non-Euclidean statistics}
\subsection{The Fr{\'e}chet mean}\label{Frechet}
When using a non-Euclidean distance $d()$ we must define what is meant by a `mean 
covariance matrix'. Consider a probability distribution for a $k \times k$ 
covariance matrix $S$ on a Riemannian metric space with 
density $f(S)$. The Fr{\'e}chet (1948) mean 
$\Sigma$ is 
defined as
$$\Sigma = \mathop{{\rm arg}\inf_{\Sigma}}  \; \;  \frac{1}{2} \int d( S , \Sigma )^2 f(S) dS , $$
and is also known as the Karcher mean (Karcher, 1977). The Fr{\'e}chet 
mean need not be unique in general, although for many distributions it will be.
Provided the distribution is supported only on the geodesic ball of radius $r$, 
such that the geodesic ball of radius $2r$ is regular (i.e. supremum of sectional curvatures 
is less than $(\pi/(2r))^2$), then the Fr{\'e}chet mean $\Sigma$ is unique 
(Le, 1995).  The support to ensure uniqueness can be very large. For example, 
for Euclidean spaces (with sectional curvature zero), or for non-Euclidean spaces with 
negative sectional curvature, the Fr{\'e}chet mean is always unique. 

If we have a sample $S_1,\ldots,S_N$ of i.i.d. observations available then the 
sample Fr\'{e}chet mean is 
calculated by finding 
$$\hat \Sigma = \mathop{{\rm arg}\inf_{\Sigma}} \sum_{i=1}^N d( S_i, \Sigma)^2 . $$
Uniqueness of the sample Fr\'{e}chet mean can also be determined from 
the result of Le (1995). 

\subsection{Distances between covariance matrices}
We now consider specific 
choices of distances in order to provide estimates of a mean from the sample of $N$ covariance 
matrices. To ensure the positive definiteness of the 
covariance matrices, a reparameterization can be used such as $S_i=Q_iQ_i^T$, where $Q_i \in {R}^{3\times 3}$. For example, $Q_i=chol(S_i)$ is the \textit{Cholesky decomposition}, where $Q_i$ is lower triangular with positive diagonal elements. Note that $Q_i$ and any rotation and reflection of it $Q_iR_i$ (${R_i \in O(3)}$) can result in the same $S_i$, i.e. $S_i=Q_iQ_i^T=Q_iR_i(Q_iR_i)^T, i=1, ..., N$.

In applications there are several choices of distances between covariance matrices that one could consider, 
for example see Table \ref{TAB1}. 

\begin{table}[htbp]
\begin{center}
\begin{tabular}{lccc}
Name & Notation & Form & Estimator\\
\hline
 Euclidean  &
$d_E(S_1,S_2)$ & $\| S_1 - S_2 \| $ & $\hat\Sigma_E$\\
Log-Euclidean  &
$ d_L(S_1, S_2)$ &  $\|  \log(S_1) - \log(S_2)  \| $ & $\hat\Sigma_L$\\
Riemannian  &
$ d_R(S_1,S_2)$ &  $\| \log (S_1^{-1/2}S_2 S_1^{-1/2}) \|$ & $\hat\Sigma_R$\\
Cholesky 
& $d_C(S_1,S_2)$ &  $\| {\rm chol}(S_1) - {\rm chol}(S_2) \| $ & $\hat\Sigma_C$\\
Root Euclidean &
$d_{H}(S_1, S_2)$ &  $  \|  S_1^{1/2}  - S_2^{1/2}  \| $ & $\hat\Sigma_H$\\ 
Procrustes size-and-shape
& $d_{S}(S_1, S_2)$ &  $\mathop{\inf_{R \in O(k)}} 
 \|  {\rm chol}(S_1)  - {\rm chol}(S_2) R  \| $ & $\hat\Sigma_S$\\
Full Procrustes shape &
$ d_F(S_1, S_2)$ & $\mathop{\inf_{R \in O(k), \beta \in \mathbf R}} 
 \left\|  \frac{ {\rm chol}(S_1) } { \| {\rm chol}(S_1)  \| }
 -  \beta {\rm chol}(S_2) R  \right\| $ & $\hat\Sigma_F$\\
Power Euclidean &
$d_{A}(S_1, S_2)$ &  $ \frac{1}{\alpha} \|  S_1^{\alpha}  - S_2^{\alpha}  \| $ & $\hat\Sigma_A$\\ 
\hline
\end{tabular}
\end{center}
\caption{{\it Some distances between covariance matrices and notation for the corresponding 
Fr\'echet mean estimators.}}\label{TAB1}
\end{table}

Estimators $\hat \Sigma_E, \hat \Sigma_C, \hat\Sigma_H, \hat \Sigma_L, \hat \Sigma_A$ given in Table \ref{TAB1} 
are straightforward to compute 
using arithmetic averages. 
Note that $d_S$ is obtained by
optimal rotation/reflection of $chol(S_2)$ onto $chol(S_1)$ using ordinary Procrustes analysis. 
The Procrustes based estimators $\hat \Sigma_S, \hat \Sigma_F$ involve the use of the Generalized Procrustes 
Algorithm, which works well in practice (see Dryden et al., 2009). 
The Riemannian metric estimator $\hat\Sigma_R$ 
uses a gradient descent 
algorithm which is guaranteed to converge (e.g. see Pennec et al, 2006). In practice it is 
similar to the log-Euclidean estimator $\hat\Sigma_L$ (Arsigny et al., 2007).

We briefly summarize some of the properties of the distances. 
All these distances are invariant under simultaneous rotation and reflection of $S_1$ and $S_2$, i.e. 
the distances are unchanged by replacing both $S_i$ by $V S_i V^T, \; V \in O(k), i=1,2$. 
 Metrics $d_L(),d_R(),d_F()$ are invariant under simultaneous scaling of 
$S_i, i=1,2$, i.e. replacing both $S_i$ by $\beta S_i$. Metric $d_R()$ is also affine invariant, 
i.e. 
the distances are unchanged by replacing both $S_i$ by $A S_i A^T, i=1,2$ where $A$ is a
general $k \times k$ full rank matrix.  
Metrics $d_L(),d_R()$ have the property that 
$d( A , I_k) = d(A^{-1}, I_k)$,
where $I_k$ is the $k \times k$ identity matrix, and 
$d_L(),d_R(),d_F()$ are not valid for comparing rank deficient covariance matrices. 
Finally, there are problems with extrapolation with metric $d_E()$: extrapolate too far and the 
matrices are no longer positive semi-definite (Arsigny et al., 2007).

An alternative anisotropy measure to FA in (\ref{FA}) is to use the full Procrustes shape distance to isotropy where
\begin{eqnarray*}
PA & = & \sqrt{ \frac{k}{k-1} } d_{F}( I_k , S)  = \left\{ \frac{k}{k-1} \sum_{i=1}^k (\sqrt \lambda_i - \overline { \sqrt \lambda } )^2 / 
\sum_{i=1}^k \lambda_i \right\}^{1/2} ,
\end{eqnarray*}
where $ \overline {\sqrt \lambda } = \frac{1}{k} \sum \sqrt{\lambda_i} $. 
We include the scale factor when defining the
Procrustes Anisotropy (PA), and so $ 0 \le PA \le 1$, with $PA = 0$ indicating isotropy, 
and $PA \approx 1$ indicating  a very strong principal axis. 
Another anisotropy measure based on metrics $d_L$ or $d_R$ is the geodesic anisotropy 
$$
GA  =  \left\{  \sum_{i=1}^k (\log \lambda_i - \overline {\log \lambda} )^2 \right\}^{1/2} , 
$$
where $0 \le GA < \infty$ (Arsigny et al., 2007), which has been used in 
diffusion tensor analysis in medical imaging with $k = 3$. Alternatively one could
consider ${\rm tanh}(GA)$ (Batchelor et al., 2005) which is on the scale $[0,1)$.

In some applications covariance matrices are close to being deficient in
rank. For example when $FA$ or $PA$ are equal to $1$ then the covariance matrix is 
of rank $1$. The Procrustes metrics can easily deal with deficient rank matrices, which is 
a strong advantage of the approach.

\section{Interpolation methods}
\subsection{Weighted Generalised Procrustes Analysis}
Frequently in diffusion tensor imaging it is of interest to interpolate between sets of tensors. 
The weighted Fr\'{e}chet sample mean of $S_1$, ..., $S_N$ at $N$ voxels with a certain distance function $d()$ is defined by:
\begin{eqnarray}
 \bar{S}&=&\arg \inf_{S} \sum \limits_{i=1}^{N} w_i d(S_i,S)^2,
\end{eqnarray}
where the weights $w_i$ are proportional to a function of the Euclidean distance between locations of the tensors (voxels), $0\leq w_i\leq 1$ and $\sum _{i=1}^{N}w_i=1$.

We choose $d_S$ for the distance and then Weighted Generalized Procrustes analysis (WGPA) is proposed to obtain the weighted mean of $S_1$, ..., $S_N$. The objective of WGPA under rotation and reflection is to minimise a sum of weighted squared Euclidean norms $S_{WGPA}$ which is given by
\begin{eqnarray}
S_{WGPA}(S_1,...,S_N)=\inf_{R_1,...,R_N} \sum \limits_{i=1}^{N} w_i\parallel Q_iR_i-\sum \limits_{j=1}^{n} w_jQ_jR_j \parallel^2\cr
=\inf_{R_1,...,R_N} \sum \limits_{i=1}^{N} w_i\parallel (1-w_i)Q_iR_i-\sum \limits_{j\neq i}w_jQ_jR_j\parallel^2\cr
=\inf_{R_1,...,R_N} \sum \limits_{i=1}^{n} \frac{w_i}{(1-w_i)^2}\parallel Q_iR_i-\frac{1}{(1-w_i)}\sum \limits_{j\neq i}w_jQ_jR_j\parallel^2.
\label{WGPA}
\end{eqnarray}
Let $\hat{R}_i, i=1,...,N$ be the estimates of the rotation matrices. Then, the \textit{WGPA mean tensor} is given by
\begin{equation}
\bar{S}_{WGPA}=\bar{Q}_{WGPA}\bar{Q}_{WGPA}^T, 
\end{equation}
where $\bar{Q}_{WGPA}=\sum \limits_{i=1}^{N} w_iQ_i\hat{R}_i$.
We give Algorithm \ref{alg} for estimating $\hat{R}_i, i=1,...,N$. Note that the algorithm is guaranteed to converge to a local minimum as the reduction in $S_c$ at each iteration is at least
zero. 

\begin{algorithm}[H]
\caption{Weighted Generalised Procrustes Method} 
   \begin{algorithmic}[1]
     \STATE \textbf{Initial setting:} $Q^P_i\leftarrow chol(D_i)$, $i=1,...,N$
     \STATE $S_{WGPA}$ from previous iteration: $S_{p}\leftarrow 0$
     \STATE $S_{WGPA}$ from current iteration: $S_{c}\leftarrow \sum \limits_{i=1}^{N} w_i\parallel Q_i^P-\sum \limits_{j=1}^{N}w_jQ_j^P \parallel^2$
     \WHILE{$|S_{p}-S_{c}|> \textnormal{tolerance}$}
     \FOR{$i=1$ to $N$}
        \STATE $\bar{Q}_i=\frac{1}{1-w_i}\sum \limits_{j\neq i}w_jQ^P_j$
        \STATE Calculate the rotation matrix $R_i$ which minimises $\parallel \bar{Q}_i-Q^P_iR_i\parallel$ with partial ordinary Procrustes analysis
        \STATE $Q^P_i \leftarrow Q^P_iR_i$
     \ENDFOR
        \STATE $S_{p}\leftarrow S_{c}$
        \STATE $S_{c} \leftarrow \sum \limits_{i=1}^{N} w_i\parallel Q_i^P-\sum \limits_{j=1}^{N}w_jQ_j^P \parallel^2$        
     \ENDWHILE
     \STATE $\bar{Q}_{WGPA} \leftarrow \sum \limits_{i=1}^N w_iQ^P_i$
     \RETURN $\bar{Q}_{WGPA}$
    \end{algorithmic}
\label{alg}
\end{algorithm}

\subsection{Regularization}
In medical image analysis a noisy tensor field may be available and so we wish to 
carry out regularization. For example, consider a grid of tensors $S_1,\ldots,S_n$ at voxels $x_1,\ldots,x_n$ 
and we wish to predict the tensor at a new site $x$. We could use the weighted penalized predictor obtained 
by minimizing, with respect to $\Sigma$, 
$$\hat\Sigma_{\beta,\omega}(\lambda) = \sum_{i=1}^n w_i {\rm dist}(S_i,\Sigma)^\beta + \lambda {\rm dist}(\Sigma,\mu)^\omega$$
where the weights $w_i \ge 0, \sum w_i = 1$ are functions of the distance to the new site, $\lambda>0$ is 
a regularization parameter, and 
$\mu$ is a reference matrix, such as the identity matrix, zero matrix or an overall average.  
For example we could use $w_i \propto \exp\{ - \gamma \| x- x_i| \|^2 \}, i=1,\ldots,n$. 

Consider now smoothing across an image at the voxels $x_1,\ldots,x_n$, and so we need to minimize, 
with respect to $\Sigma_j, j=1,\ldots,n$,
$$ \sum_{j=1}^n \sum_{i=1}^n w_{ij} {\rm dist}(S_i,\Sigma_j)^\beta  + \lambda \sum_{j=1}^n 
{\rm dist}(\Sigma_j, \mu )^\omega , $$
and $w_{ij}$ is the weight as a function of the distance between sites $i$ and $j$. 
Note $(\beta,\omega) = (2,0)$ gives the weighted Fr\'echet mean, if $(\beta,\omega) = (\beta,0)$ 
we have a type of M-estimator (Kent, 1992; Dryden and Mardia, 1998, p298), if
 $(\beta,\omega) = (1,0)$ we have the geometric 
median (Fletcher and Joshi, 2009), if $(\beta,\omega) = (2,2)$ non-Euclidean type of ridge-regression, 
and if $(\beta,\omega) = (2,1)$ a non-Euclidean type of LASSO (see Tibshirani, 1996). 
Note that for the power metric (and Euclidean and
square root) the space is Euclidean, and so using this procedure is relatively 
straightforward in this case.

\section{Applications}
\subsection{Anisotropy of diffusion tensors}
We consider anisotropy of estimated diffusion tensors in the brain obtained 
from diffusion weighted images (see Dryden et al., 2009). In Figure \ref{DTI} we see a  coronal view of the brain, and the corpus callosum and cingulum 
can be seen.  

\begin{figure}[htbp]
\begin{center}
\includegraphics[width=6cm,angle=270]{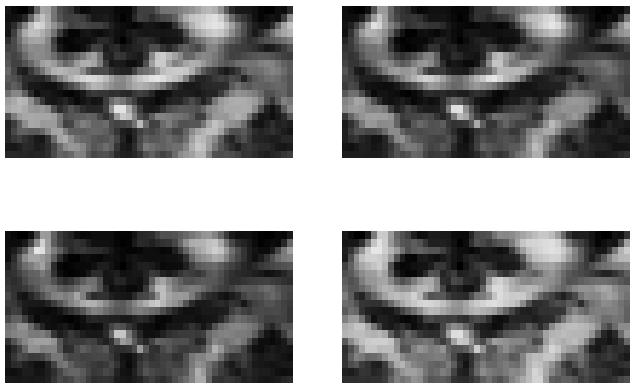}
\end{center}
\caption{\it The anisotropy measures (top left) FA, (top right) PA, (bottom left) GA and (bottom right) $tanh(GA)$ }
\label{DTI}
\end{figure}

At first sight all three anisotropy measures appear broadly similar. 
However, the PA image offers more contrast than the FA image in the highly anisotropic region - the corpus callosum. Also, the GA image has rather fewer brighter areas than PA or FA. The plot of $tanh(GA)$ is most different from the others, with much fewer dark areas. 
Due to the improved contrast 
we believe PA is slightly  
preferable in this example.

\begin{figure}[htbp]
\begin{center}
\includegraphics[width=10cm]{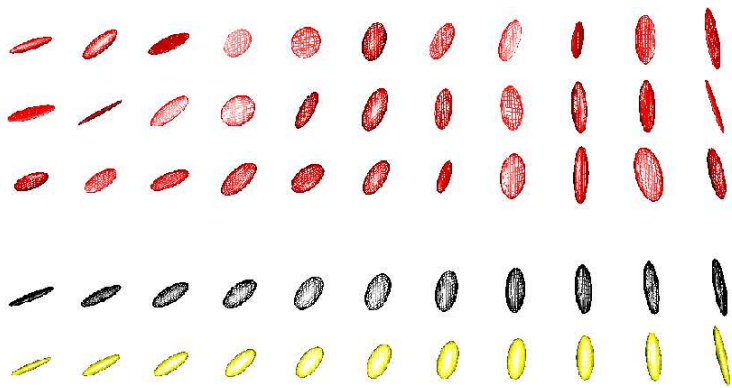}
\end{center}
\caption{
\it Principal geodesic analysis for covariance matrices. 
The true geodesic path is given in the penultimate row (black).  
We then add noise in the three initial rows (red). Then we
estimate the mean and find the first principal component (yellow), displayed in the 
bottom row.
}
\label{PCAnoisy}
\end{figure}

\subsection{Principal geodesics of covariance matrices}
We consider now an example estimating the principal geodesics of the 
covariance matrices $S_1,\ldots,S_n$ using the Procrustes size-and-shape 
metric $d_S$ (see Dryden et al., 2009). Huckeman et al. (2009) discuss 
geodesic principal components analysis in Riemannian manifolds in depth. We consider an 
approximate procedure where the principal geodesics are estimated by principal components
analysis of the tangent space co-ordinates. 
In Figure \ref{PCAnoisy}, we consider a 
true geodesic path (black) and evaluate $11$ equally spaced covariance 
matrices along this path. We then add i.i.d. Gaussian noise in the tangent space 
for three separate 
realisations of noisy paths (in red). 
The overall 
mean $\hat\Sigma_S$ is computed based on all the data ($n=33$), 
and then the Procrustes size-and-shape 
tangent space co-ordinates are obtained based on the Cholesky decompositions 
of the covariance matrices. The first principal component loadings are 
computed and projected back to give an estimated minimal 
geodesic in the covariance matrix space. We plot this path in yellow 
by displaying 
$11$ covariance matrices along the path. It can be seen that the estimated principal 
geodesic is very similar to the true geodesic path here. Other extensions include curve fitting through 
paths of covariance matrices using polynomials and geodesics (e.g. see Evans et al., 2009, for some
examples of shape curves).

\subsection{Interpolation}\label{interp2}
A tensor field from a healthy human brain has been smoothed and interpolated (with 2 interpolations between each pair of original voxels). The Fractional Anisotropy (FA) maps from the processed tensors are shown in Figure \ref{fig com}. Obviously, the FA map from the processed tensor data is much smoother than the one without processing. The feature that the cingulum is distinct from the corpus callosum is clearer in the anisotropy map from the processed data than those without processing in Figure \ref{fig com}.
\begin{figure}[htbp]
 \centering
 \includegraphics[height=1.8in]{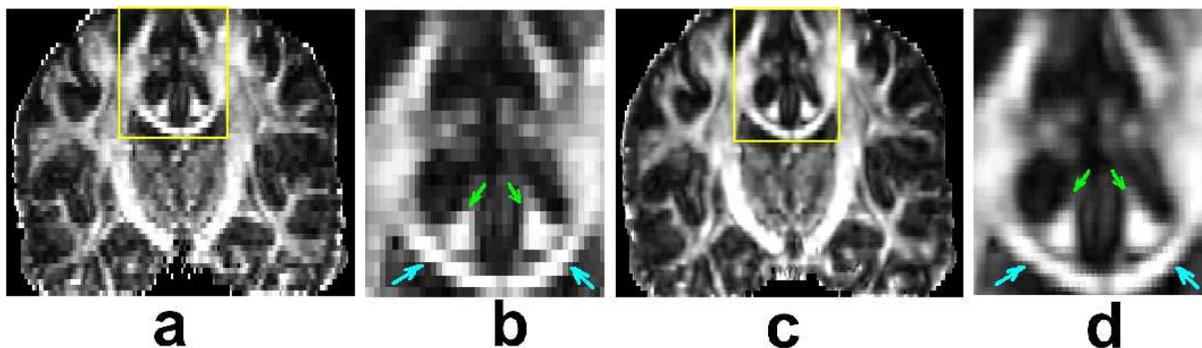}
\vspace{-0.2in}
\caption{\it Smoothing and interpolation of the diffusion tensor data from human brain. a: FA map from Bayesian tensor field. c: FA map from processed tensor field. b and d: Zoomed inset regions. Green arrows: the cingulum. Light blue arrows: the corpus callosum.}
\label{fig com} 
\end{figure}

\subsection{Tractography}
As a final application we give some initial results of fibre tractographies of the brain stem in a healthy human 
in Figure~\ref{fig track}. It is of great interest to study the white matter fibre tracts in the brain in 
order to explore connectivity between different parts, both in healthy and patient brains.  
From different seed points in the brain stem, 
white matter fibres are tracked by following interpolated paths of principal directions 
from diffusion tensors. 
Tractography from the 
WGPA processed tensor field is different from the other methods, and work is currently underway to assess whether WGPA is preferable.

\begin{figure}[htbp]
\includegraphics[height=1.6in]{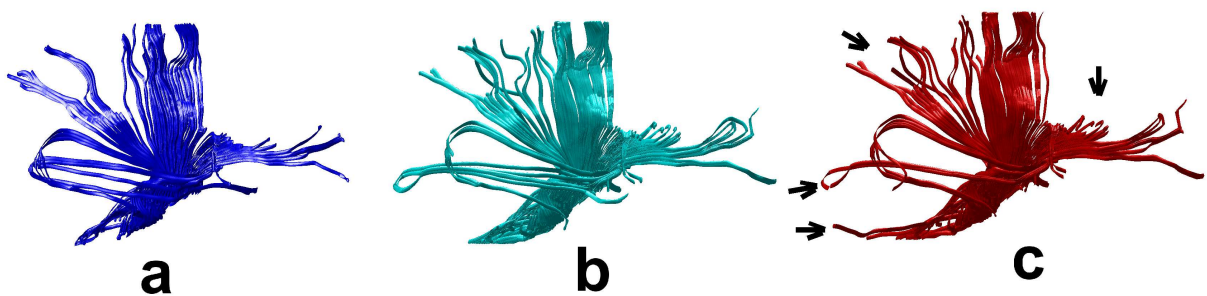}
\vspace{-0.2in}
 \caption{\it Fibre tractograhpies using the Bayesian estimates (a), Euclidean smoothing (b) and WGPA smoothing (c). Black arrows point out some obvious differences of the WGPA tracts compared with other methods.}
\label{fig track} 
\end{figure}

\section{Conclusions}
Methodology for estimation and inference in the 
space of covariance matrices has application in 
many areas, including diffusion tensor imaging, structural tensor analysis 
in computer vision, and modelling longitudinal data with Bayesian and random effect models.  
There are many choices of metric available, each with its advantages. The particular choice of what is best will 
depend on the particular application. 
The use of the Procrustes size-and-shape metric $d_S$ is particularly 
appropriate when the covariance matrices are close to being deficient in rank.

\end{ISItext}

\ISIsubtitleB{REFERENCES (R\'EFERENCES)}

\begin{ISItext}
\begin{small}
\begin{description}

\item Alexander, D.~C. (2005).
\newblock Multiple-fiber reconstruction algorithms for diffusion {MRI}.
\newblock {\em Ann NY Acad Sci}, 1064:113--133.

\item Arsigny, V., Fillard, P., Pennec, A.  and Ayache, N. (2007). 
{Geometric Means in a Novel Vector Space Structure on Symmetric Positive-Definite Matrices}, 
{\em SIAM Journal on Matrix Analysis and Applications}, {\bf 29}, 328--347.

\item
Basser, P.~J., Mattiello, J., and Le~Bihan, D. (1994).
 Estimation of the effective self-diffusion tensor from the {NMR} spin
  echo.
 {\em J Magn Reson B.}, {\bf 103}, 247--254.

\item Batchelor P.G, Moakher M, Atkinson D, Calamante F, Connelly A. (2005). A rigorous framework for diffusion tensor calculus.
{\it Magn. Reson. Med.}, {\bf 53}, 221--225.

\item
Dryden, I.~L., Koloydenko, A., and Zhou, D. (2009).
 Non-Euclidean statistics for covariance matrices, with applications
  to diffusion tensor imaging.
 {\em Annals of Applied Statistics}.
 To appear.

\item Dryden, I.~L. and Mardia, K.~V. (1998).
\newblock {\em Statistical Shape Analysis}.
\newblock Wiley, Chichester.

\item
Evans, K., Dryden, I.~L., and Le, H. (2009).
 Shape curves and geodesic modelling.
 Technical report, Division of Statistics, University of Nottingham.
 Submitted for publication.

\item
Fillard, P., Arsigny, V., Pennec, X., and Ayache, N. (2007).
 Clinical {DT-MRI} estimation, smoothing and fiber tracking with
  log-{E}uclidean metrics.
 {\em IEEE Transactions on Medical Imaging}, {\bf 26}, 1472--1482.

\item Fletcher, P.T.,  Venkatasubramanian, S. and Joshi, S. (2009). 
The geometric median on Riemannian manifolds with application to robust atlas estimation. 
45, S143-S152. 

\item Fr{\'e}chet, M. (1948).
\newblock Les \'el\'ements al\'eatoires de nature quelconque dans un espace
  distanci\'e.
\newblock {\em Ann. Inst. H. Poincar\'e}, 10:215--310.

\item Huckemann, S., Hotz, T. and Munk, A. (2009). Intrinsic Shape Analysis:
Geodesic Principal Component Analysis for Riemannian Manifolds Modulo
Lie Group Actions. Discussion paper, Statistica Sinica, to appear.

\item Karcher, H. (1977).
\newblock Riemannian center of mass and mollifier smoothing.
\newblock {\em Comm. Pure Appl. Math.}, 30(5):509--541.

\item Kent, J.~T. (1992).
\newblock New directions in shape analysis.
\newblock In Mardia, K.~V., editor, {\em The Art of Statistical Science}, pages
  115--127. Wiley, Chichester.

\item Le, H.-L. (1995).
\newblock Mean size-and-shapes and mean shapes: a geometric point of view.
\newblock {\em Advances in Applied Probability}, 27:44--55.

\item Pennec, X., Fillard, P. and Ayache, N. (2006). 
 A Riemannian Framework for Tensor Computing, {\em Int. J. Comput. Vision},
 {\bf 66},  41--66.

\item Tibshirani, R. (1996). Regression shrinkage and selection via the lasso.  {\em J. Royal. Statist. Soc B.}, Vol. 58,  26-7-288).

\item Zhou, D., Dryden, I.~L., Koloydenko, A., and Bai, L. (2008).
\newblock A {B}ayesian method with reparameterisation for diffusion tensor
  imaging.
\newblock In Reinhardt, J.~M. and Pluim, J. P.~W., editors, {\em Proceedings,
  SPIE conference. Medical Imaging 2008: Image Processing}, page 69142J.

\end{description}
\end{small} 
\end{ISItext}

\end{document}